\pdfoutput=1
\documentclass[10pt,pra,aps,superscriptaddress,nofootinbib,twocolumn,longbibliography]{revtex4-1}
\usepackage{amsmath,bbm}
\usepackage{amsthm}
\usepackage{amsmath}
\usepackage{latexsym}
\usepackage{amssymb}
\usepackage{float}
\usepackage{graphicx}           
\usepackage{color}
\usepackage{xcolor}
\usepackage{mathpazo}
\usepackage{comment}
\usepackage{enumitem}
\usepackage{multirow}
\usepackage{setspace}
\usepackage{subcaption}
\usepackage[colorlinks=true,linkcolor=blue,citecolor=magenta,urlcolor=blue]{hyperref}

\newcommand{\be}{\begin{equation}}
\newcommand{\ee}{\end{equation}}
\newcommand{\bea}{\begin{eqnarray}}
\newcommand{\eea}{\end{eqnarray}}
\def\squareforqed{\hbox{\rlap{$\sqcap$}$\sqcup$}}
\def\qed{\ifmmode\squareforqed\else{\unskip\nobreak\hfil
\penalty50\hskip1em\null\nobreak\hfil\squareforqed
\parfillskip=0pt\finalhyphendemerits=0\endgraf}\fi}
\def\endenv{\ifmmode\;\else{\unskip\nobreak\hfil
\penalty50\hskip1em\null\nobreak\hfil\;
\parfillskip=0pt\finalhyphendemerits=0\endgraf}\fi}

\newcommand{\tr}{\text{Tr}}
\newcommand{\I}{\mathbbm{1}}

\newcommand{\ket}[1]{|#1\rangle}
\newcommand{\bra}[1]{\langle#1|}

\newcommand{\la}{\langle}
\newcommand{\ra}{\rangle}

\makeatletter
\newtheorem*{rep@theorem}{\rep@title}
\newcommand{\newreptheorem}[2]{%
\newenvironment{rep#1}[1]{%
 \def\rep@title{#2 \ref{##1}}%
 \begin{rep@theorem}}%
 {\end{rep@theorem}}}
\makeatother

\newtheorem{thm}{Theorem}
\newreptheorem{thm}{Theorem}
\newtheorem{lemma}{Lemma}

\newtheorem{obs}{Observation}

\begin{document}

\title{ Usefulness of $d\otimes d$ dimensional pure entangled states for antidiscrimination of quantum measurements when $d$ is even}

\author{Satyaki Manna}
\email{mannasatyaki@gmail.com}
\affiliation{School of Physics, Indian Institute of Science Education and Research Thiruvananthapuram, Kerala 695551, India}
\affiliation{Department of Physics, School of Basic Sciences, Indian Institute of Technology Bhubaneswar, Odisha 752050, India}

\begin{abstract}
     The usefulness of entanglement in quantum channel discrimination is a central topic in quantum information theory. A landmark result in this direction was established by
    Piani and Watrous [$Phys. Rev. Lett.102, 250501 (2009)$] who proved that all entangled states are useful for discrimination of quantum channels. We pose the same question in the context of antidiscrimination of quantum channels. We partially answer this by showing that for every pure entangled state of $\mathbbm{C}^d\otimes\mathbbm{C}^d$ (with even $d$), there exist three projective measurements which are antidiscriminable (but not discriminable) with that input state but those three measurements are not antidiscriminable with the product probe.
\end{abstract}

\maketitle


\section{Introduction}
Entanglement is one of the most fascinating features of quantum theory. As a nonlocal resource, it serves as the fundamental cornerstone of quantum information theory by playing a pivotal role in various information-theoretic tasks such as quantum teleportation \cite{teleportation,teleport}, dense coding \cite{dc,densecoding}, quantum cryptography \cite{crypto}, quantum error correction \cite{ekert,ec}, quantum information \cite{nielsen2010quantum,10.5555/3240076,masanes,sarkar} and quantum communication \cite{gisin2007quantum, bhaumik2025}.

Another important aspect of entanglement was discovered when Kitaev \cite{AYuKitaev_1997} demonstrated that an entanglement-assisted probe–ancilla system can sometimes offer an advantage in discriminating quantum channels. Over the years, the advantage provided by entanglement over product probe systems in channel discrimination task has been extensively studied \cite{sacchi,sacchi2,Piani,Bae,Datta_2021,manna,uola}. However, most applications of entanglement have focused on specific classes of entangled states, even when the system dimension is fixed, with only a few notable exceptions such as \cite{Piani,Bae}. In Ref. \cite{Piani}, the authors showed that every entangled state is useful for the discrimination of two channels. We address the analogue question of \cite{Piani} in the context of antidiscrimination of quantum channels, i.e., for every entangled state, can we find three indistinguishable channels which are antidiscriminable with that entangled state but those three channels are not antidistinguishable with the product probe? Our answer is partial; we prove the affirmative of this question for pure entangled states of $\mathbbm{C}^d\otimes\mathbbm{C}^d$ when $d$ is even. In this case, the channels we take are quantum measurements, which can be implemented as Quantum-to-Classical channels\cite{2016,10.5555/3240076}. Taking the dimension of ancilla the same as the system can be motivated from a previous result of measurement antidistinguishability\cite{manna}. We not only prove the existence of such measurements but also provide the explicit construction of those for any pure entangled states of $\mathbbm{C}^d\otimes\mathbbm{C}^d$.

 The notion of antidiscrimination captures the possibility of ruling out certain alternatives in a
quantum experiment without identifying the actual outcome. This can be labeled as a weaker version of discrimination, which refers to identifying which specific process has occurred from a set of all
possibilities. In addition to its application to foundational insights into the reality of quantum states \cite{pbr,chaturvedi2021,Chaturvedi2020,ray2024,ray2025,bhowmik2022,branciard,leifer}, both antidiscrimination and discrimination have been  analyzed in the context of quantum information \cite{uola} and quantum communication tasks \cite{barett,manna2,pandit2025}. Although discrimination of quantum states \cite{Hellstrom,benett,vedral,walgate,watrous,paul,schmid,bandyo,halder,ghosh,ghosh_s} and channels \cite{Piani,manna,harrow,bsxv-q9x7,manna2025,duan,Bae,acin,acin2,watrous} is a well developed field, antidiscrimination tasks is a less explored area till now. Despite being fewer in number, there is a considerable body of work addressing the antidistinguishability of quantum
states \cite{Caves02,Heinosaari_2018,barett,con_ex,Johnston2025,yao2025}. However, the study of antidistinguishability for quantum channels remains almost non-existent \cite{manna3,PRXQuantum,ji2025barycentricbounds}. 

In this work, our main goal is to show the usefulness of all pure entangled states of $\mathbbm{C}^d\otimes\mathbbm{C}^d$ for the antidiscrimination of quantum measurements when $d$ is even. At first, we formalize the notion of antidiscrimination of quantum measurement for single system probe and entangled probe. In the entanglement assisted scenario, the question of the dimension of the ancillary system is addressed from a previously proven result of \cite{manna}. This gives the reason to consider qudit-qudit entangled states in order to find antidistinguishability of qudit measurements. In these protocols, we do not have access to the post-measurement states. Then we dive into our main result, which is divided into three theorems. Firstly, we prove the usefulness of qubit-qubit entangled states. Subsequently, we prove the same inference for the entangled states of dimension of multiple of four and then for the entangled states of the rest even dimensions.

We start by giving a brief review of the antidiscrimination of quantum states because some of the results will be necessary in the course of the paper. Consequently, we move to the formulation of antidistinguishability of the measurements with single system and entangled probe. Then we present our claim in the section \ref{res}.

\section{Antidiscrimination of quantum states}
 Assume, we are given $n$ previously known quantum states $\{\rho_k\}_{k=1}^n$ associated with the positive numbers $\{q_k\}_{k=1}^n$. In general, $\{q_k\}_{k=1}^n$ may not form probability distribution. The experimentalist knows the set of the states as well as their associated positive numbers. One can consider a black box where any one of the states from the set is supplied from a referee. The experimentalist needs to infer which of the states is not supplied.
 Antidistinguishability of $n$ quantum states $\{\rho_k\}_{k=1}^n$ is a function of a set of positive numbers $\{q_k\}_k$, which is defined as,
\bea  
\mathcal{AS}[\{\rho_k\}_k,\{q_k\}_k] &= & \max_{M=\{M_a\}_a}\Bigg\{ \sum_{k,a} q_k  p(a\neq k|\rho_k,M)\Bigg\}.\nonumber\\
\eea
 We are taking optimization over the $n$-outcome positive operator valued measurement $\{M_a\}_{a=1}^n$ such that the above quantity attains its maximum value. 
As $\sum_a (p(a\neq k|\rho_k,M)+p(a=k|\rho_k,M))=1$ for any $\{M_a\}_a$ , the above expression becomes,
\bea\label{pA}
AS[\{\rho_k\}_k,\{q_k\}_k] &=& \sum_k q_k - \min_{\{M_a\}_a}\Bigg\{\sum_{k}q_k \ p(a=k|\rho_k,M)\Bigg\} \nonumber \\
&=& \sum_k q_k - \min_{\{M_a\}_a}\left\{\sum_{k}q_k \ \tr(\rho_k M_k)\right\} .
\eea 

 The second part of the last line is the direct implication of Born rule.

If three states $\rho_1, \rho_2$ and $\rho_3$ are perfectly antidistinguishable, the above equation reads,
$AS[\{\rho_k\}_k,\{q_k\}_k] = \sum_k q_k$.
Note, we do not impose $\sum_k q_k=1$, i.e., $\{q_k\}_k$ may not be a probability distribution, on purpose. In the next section, we will see that the antidistinguishability of the measurements will be related to the problem of antidistinguishability of the states such that the associated coefficients $\{q_k\}_k$ may not form a probability distribution. 

The necessary and sufficient conditions \cite{Caves02} for three pure quantum states which are pairwise non-orthogonal to be perfectly antidistinguishable, i.e.,
\be \label{ADSpsi123}
\mathcal{AS}[\{\psi_1,\psi_2,\psi_3\},\{q_1,q_2,q_3\}] = \sum_k q_k,
\ee 
are following:
\begin{subequations}\label{condAS}
\be
    x_1 + x_2 + x_3 < 1 
\ee
\be 
(x_1+x_2+x_3-1)^2 \geqslant 4x_1x_2x_3 ,
\ee 
\text{  where $x_1=|\la \psi_1|\psi_2\ra|^2, x_2=|\la \psi_1|\psi_3\ra|^2, x_3=|\la \psi_2|\psi_3\ra|^2$.}\\
\end{subequations}

The ref. \cite{barett} states, the sufficient condition of antidiscrimination of three pure states, i.e., $\ket{\psi_1},\ket{\psi_2},\ket{\psi_3}$ is $x_1,x_2,x_3\leqslant 1/4$, where $x_1=|\la \psi_1|\psi_2\ra|^2, x_2=|\la \psi_1|\psi_3\ra|^2, x_3=|\la \psi_2|\psi_3\ra|^2$. In our proofs, we mainly use this condition. 
\begin{lemma}\label{suff_con}
    The necessary and sufficient condition for a set of pure qubit states $\{\rho_i\}_{i=1}^n$ to be anti-distinguishable is the following:\\
    $\sum_i\mu_i\rho_i=\I$, for some positive real numbers $\mu_i$.
\end{lemma}
\begin{proof}
See proposition $6$ of \cite{Heinosaari_2018}.
\end{proof}

\section{Antidiscrimination of quantum measurements}
Quantum measurement having $f$ distinct outcomes is defined by a set of operators,
$\mathcal{M} :=\{G_a\}_{a=1}^f$,  
such that $\sum_{a=1}^f G_a^\dagger G_a = \I$. When this measurement is performed on a quantum state $\rho$,
the probability of obtaining outcome $a$ is given by Born rule, which is-
$p(a|\rho) = \tr(\rho G^\dagger_a G_a)$.
Projective measurement is a particular case where $G_a$ are projectors.

We consider a priori known set of $l$ measurements acting on $d$-dimensional quantum states, each having $f$ outcomes, defined by $\{G_{a|x}\}_{a,x}$, where $x\in\{1,\cdots,l\}$ denotes the measurements and $a\in\{1,\cdots,f\}$ denotes the outcomes. The measurements are sampled from a probability distribution $\{p_x\}_x$, i.e., $p_x > 0, \sum_x p_x =1$. Note that, in general, $d$ may not be same as $f$. We define the positive operators, commonly known by POVM elements, as
$\mathcal{M}_{a|x} = G^\dagger_{a|x}G_{a|x}$ ,
which is convenient to use because we do not consider the post-measurement states here. To antidistinguish these $l$ measurements $\{G_{a|x}\}_{a,x}$, the measurement device is given a known quantum state, which can be single or entangled, and the device carries out one of these $l$ measurements. Measurements and this initial quantum state belong to $\mathbbm{C}^d$. Now we are going to discuss two types of antidiscrimination protocols depending on the input state.\\

\subsection{Antidiscrimination with single system probe}
At first, we start with a single system as the initial state and only have the classical output of the measurement device (See FIG 1a of \cite{manna}). Note, we do not have access to the post-measurement state. In this scenario, we will get an outcome with the classical probability which is denoted by $p(a|x,\rho)$. Upon receiving this classical output, in general, one can perform a classical post-processing defined by $p(z|a)$, where $z\in\{1,\cdots,l\}$ and $\forall a$, $\sum_z p(z|a)=1.$ Post-processing protocol acts on $a$ and returns output $z$ for a given $a$ such that $z$ is not the guess of the measurement, which implies $z\neq x$. Hence, the antidistinguishability of the set of quantum measurements $\{G_{a|x}\}_{a,x}$ sampled from probability distribution $\{p_x\}_x$ with single system, denoted by `AMS', is given by  
\bea\label{AMS}
&&AMS\left[\left\{G_{a|x}\right\}_{a,x},\left\{p_x\right\}_x\right] \nonumber \\ 
 & = & \max_{\rho,p(z|a)} \sum_{x,a,z} p_x p\left(z\neq x|a\right) p\left(a|x,\rho\right) \nonumber\\
  &=&  1 - \min_{\rho} \sum_{a,x} p_x p(x|a) p\left(a|x,\rho\right) \nonumber\\ 
 &=&  1 - \min_{\rho} \sum_{a} \min_x \left\{p_x p\left(a=x|x,\rho\right)\right\} \nonumber\\ 
 & = & 1 - \min_{\rho} \sum_{a} \min_x\left\{p_x\tr\left(\rho \mathcal{M}_{a|x}\right)\right\},
\eea
 where $\mathcal{M}_{a|x}=G_{a|x}^\dagger G_{a|x}$.
The fourth line is written from the fact that $\sum_x p(x|a)=1$.

\begin{lemma}\label{lem2}
A set of measurements $\{G_{a|x}\}_{a=1}^{f}$ $(x = 1, 2, \ldots, l)$ is perfectly
antidistinguishable in this scenario, i.e.,
$AMS\left(\{G_{a|x}\}_{a,x}, \{p_x\}_x\right)=1$, if and
only if there exists a probe state $\sigma$ such that for every
$a \in \{1,2,\ldots,f\}$, there exists an $x \in \{1,2,\ldots,l\}$
satisfying $G_{a|x}\sigma=\sigma G_{a|x}^{\dagger}=0$.    
\end{lemma}

\begin{proof}
Without loss of generality, we may assume that $p_x>0$ for all
$x\in\{1,2,\ldots,l\}$. Using Eq.~(5),
\[
AMS\left(\{G_{a|x}\}_{a,x},\{p_x\}_x\right)=1
\]
if and only if
\begin{equation}\label{eq6}
\min_{\sigma}\sum_a\min_x
p_x\tr\left(\sigma G_{a|x}^{\dagger}G_{a|x}\right)=0.
\tag{6}
\end{equation}

The above equation is equivalent to there exists a probe state $\sigma$ such that
\[
\min_x
p_x\tr\left(\sigma G_{a|x}^{\dagger}G_{a|x}\right)=0,
\qquad
\forall a\in\{1,2,\ldots,f\}.
\]

Equivalently, for every $a\in\{1,2,\ldots,f\}$, there exists an
$x\in\{1,2,\ldots,l\}$ such that
\[
\tr\left(\sigma G_{a|x}^{\dagger}G_{a|x}\right)
\equiv
\tr\left(G_{a|x}\sigma G_{a|x}^{\dagger}\right)
\]
\[
\equiv
\tr\left((G_{a|x}\sigma^{1/2})(G_{a|x}\sigma^{1/2})^{\dagger}\right)
=0,
\]
i.e.,
\[
G_{a|x}\sigma=\sigma G_{a|x}^{\dagger}=0.
\]
\end{proof}

In the Results section, we consider Eq.~\eqref{eq6}. As already mentioned, since we are not concerned with the post-measurement state, it is sufficient to consider $\mathcal{M}_{a|x}$ for the classical probability. 

\subsection{Antidiscrimination with entangled probe}
In this setup, the antidistinguishability of a set of quantum measurements $\{G_{a|x}\}_{a,x}$ is studied using a known quantum bipartite state $\rho^{AB}$ shared by two observers, say, Alice and Bob (see FIG 1b in \cite{manna}). One of the measurements from the set is carried out on Alice's subsystem of the entangled system, and an outcome $a$ is obtained, which is conveyed to Bob. We define the reduced state of Bob when outcome $a$ is obtained by Alice for measurement $x$, 
\be \label{rhoBax}
\rho^B_{a|x} = \tr_A\left[\frac{\left(G_{a|x}\otimes \I\right)\rho^{AB}\left(G^\dagger_{a|x}\otimes \I\right)}{\tr\left( \rho^{AB} \left( \mathcal{M}_{a|x} \otimes \I \right) \right)}  \right],
\ee 
where $p(a|x,\rho^{AB}) = \tr \left[(\mathcal{M}_{a|x}\otimes \I)\rho^{AB} \right]>0 $ and $\mathcal{M}_{a|x}=G_{a|x}^\dagger G_{a|x}$.\\

Depending on Alice's outcome, Bob performs a measurement described by the set of POVM elements $\{N_{b|a}\}$ where $b\in\{1,\cdots,l\}$ is the outcome and $a\in\{1,\cdots,f\}$ is the choice of measurement settings, $\{N_{b|a}\}\geqslant 0, \sum_b N_{b|a}=\I$. This protocol is successful when Bob's output $b$ is not equal to Alice's input $x$. Note that since Bob can choose the best possible measurement on his side, any classical post-processing of the outcome of his measurement can be absorbed within the measurement $\{N_{b|a}\}$. The antidistinguishability of quantum measurements with entangled systems, denoted by 'AME', is written as,
\bea
&& AME\left[\left\{G_{a|x}\right\}_{a,x},\left\{p_x\right\}_x\right] \nonumber\\
&=& \max_{\rho^{AB}, \{N_{b|a}\}}\sum_{x,a}\sum_{b\neq x} p_x p\left(a|x,\rho^{AB}\right)p\left(b|a\right) \nonumber\\
&=& \max_{\rho^{AB}, \{N_{b|a}\}} \ \sum_{x,a}\sum_{b\neq x} p_x \operatorname{Tr} \left[\rho^{AB} \left( \mathcal{M}_{a|x} \otimes N_{b|a}\right)\right] \nonumber\\ 
&=& \max_{\rho^{AB}, \{N_{b|a}\}} \ \sum_{x,a} p_x \operatorname{Tr} \left[\rho^{AB} \left( \mathcal{M}_{a|x}\otimes(\I - N_{b= x|a})\right)\right] \nonumber\\
&=&
\max_{\rho^{AB}}
\Big[
\sum_{x,a}
p_x\tr\!\left[
\rho^{AB}\,
(\mathcal{M}_{a|x}\otimes \I)
\right]
\nonumber\\
&&-
\min_{\{N_{b|a}\}}
\sum_{x,a}
p_x\tr\!\left[
\rho^{AB}\,
(\mathcal{M}_{a|x}\otimes N_{x|a})
\right]
\Big]
\nonumber\\
&=& 
\max_{\rho^{AB}}\sum_a\Big(\sum_x p_xp(a|x,\rho_{AB})\nonumber\\
&&-
\min_{\{N_{b|a}\}}
\sum_{x}
p_x\,p(a|x,\rho_{AB})
\tr\!\left[
\rho_{a|x}^{B}N_{x|a}
\right]\Big).\nonumber\\ 
\eea

 The second part of the last line comes from the use of \eqref{rhoBax}. 
The term within the summation over $a$ coincides with the antidistinguishability of quantum states $\left\{\rho^{B}_{a|x} \right\}_x$ associated with the positive numbers $\{p_xp(a|x,\rho^{AB})\}_x$ introduced in \eqref{pA}. This implies,
\bea\label{AME}
&& AME\left[\left\{G_{a|x}\right\}_{a,x},\left\{p_x\right\}_x\right] \nonumber\\
&=& \max_{\rho^{AB}} \sum_a AS\left[\left\{\rho^{B}_{a|x} \right\}_x,\left\{p_x p\left(a|x,\rho^{AB}\right) \right\}_x\right].
\eea

 One can check $\sum_x p_xp(a|x,\rho^{AB})\neq 1$, though all of them are positive numbers. That is the reason of taking $\{q_k\}_k$ positive numbers but not probability distribution in the eq. \eqref{pA}. 

In the next section, we use the kind of construction where none of the sets among $\left\{\rho^{B}_{a|x} \right\}_x$ is distinguishable and none of the operators $\{N_{b|a}\}$ is null operator for all $a$.

We now present a useful lemma concerning the choice of entangled probes in this scenario. This lemma also clarifies the motivation for focusing on the usefulness of entangled states acting on $\mathbbm{C}^d \otimes \mathbbm{C}^d$.
\begin{lemma}
    For qudit measurements, it is sufficient to consider qudit-qudit entangled states in order to find AME.
\end{lemma}
\begin{proof}
    See Theorem $4$ of \cite{manna}.
\end{proof}
\begin{obs}
    If the probe is product state in this protocol, then the  scenario reduces to that with single system probe.
\end{obs}

It is needless to say, if the probe is product state, Bob's system does not change according to Alice's operation and antidiscrimination depends only on the classical outcomes.  

For more details of these scenarios, one can check the ref. \cite{manna}.

\section{Results}\label{res}
In this section, we discuss our main claim. In this section, we denote the entangled states of $\mathbbm{C}^d \otimes \mathbbm{C}^d$ by $d\otimes d$ entangled states. For any $d\otimes d$ entangled state, we prove that we can find three measurements which are antidiscriminable and these three measurements are not antidistinguishable with any single (product) system probe. As it is very difficult to find some general form of measurements for any even dimension, we first show that any qubit-qubit entangled state, our argument holds. All the measurements taken in this section are sampled from equal probability distribution.

As a first step, we want to prove the usefulness of qubit-qubit entangled states for the antidistinguishability of three qubit measurements. For that reason, let us consider three qubit projective measurements as follows:
\bea\label{R}
\{R_{a|1}\}_{a=1}^2 &=& \{\ket{\psi}\bra{\psi}, \ket{\psi^\perp}\bra{\psi^\perp}\},\nonumber\\
\{R_{a|2}\}_{a=1}^2 &=& \{\ket{\psi'}\bra{\psi'}, \ket{\psi'^\perp}\bra{\psi'^\perp}\},\nonumber\\
\{R_{a|3}\}_{a=1}^2 &=& \{\ket{\Bar{\psi}}\bra{\Bar{\psi}}, \ket{\Bar{\psi}^\perp}\bra{\Bar{\psi}^\perp}\},
\eea
where $\ket{\psi'}=\cos x \ket{\psi} + e^{\mathbbm{i}\theta}\sin x\ket{\psi^\perp}$ and $\ket{\Bar{\psi}}=\cos x \ket{\psi} - e^{\mathbbm{i}\theta}\sin x\ket{\psi^\perp}$.
\begin{thm}
 $(i)$  Any general qubit-qubit entangled state $\ket{\phi_2}=\sqrt{\lambda}\ket{\psi 1}+e^{\mathbbm{i}\theta}\sqrt{1-\lambda}\ket{\psi^\perp 2}$ can be used as a successful probe to antidiscriminate three qubit projective measurements described by \eqref{R} such that $\tan^2 x > \max\{\frac{\lambda}{1-\lambda},\frac{1-\lambda}{\lambda}\}$, where $x\in(0,\pi/2)$.\\
    
$(ii)$    These measurements of \eqref{R} are not antidiscriminable with single system.
    
\end{thm}
 Leveraging \eqref{AME}, we can write the antidistinguishability of three measurements and for the perfect antidistinguishability of these reduced qubit states, we use Lemma \ref{suff_con}. Then the necessary condition follows from the straight-forward calculation. The calculation is postponed to Appendix \ref{pr1}. 

Now we will proof our claim for the dimensions $(m)$  which are multiple of four, i.e., $m=4p$, where $p\in(1,\cdots,x)$. For this reason, consider the following three $m$-dimensional projective measurements.
\bea\label{S}
\{S_{a|1}\}_{a=1}^m &=& \{\ket{\eta_a}\bra{\eta_a}\}_{a=1}^m,
\{S_{a|2}\}_{a=1}^m = \{\ket{\psi^s_a}\bra{\psi^s_a}\}_{a=1}^m,\nonumber\\
\{S_{a|3}\}_{a=1}^m &=& \{\ket{\Bar{\psi^s_a}}\bra{\Bar{\psi^s_a}}\}_{a=1}^m,
\eea 
where,
$\ket{\psi^s_a}=\omega\ket{\eta_a}+\sqrt{1-\omega^2}\ket{\eta_{a+1}}$ for $a=\{1,3,5,\cdots,m-1\}$ and $\ket{\psi^s_a}=\sqrt{1-\omega^2}\ket{\eta_{a-1}}-\omega\ket{\eta_{a}}$ for $a=\{2,4,6,\cdots,m\}$, \\
$\ket{\Bar{\psi^s_a}}=\omega\ket{\eta_a}+\sqrt{1-\omega^2}\ket{\eta_{d+1-a}}$ for $a=\{1,\cdots,m/2\}$ \\
and $\ket{\Bar{\psi^s_a}}=\sqrt{1-\omega^2}\ket{\eta_{d+1-a}}-\omega\ket{\eta_{a}}$ for $a=\{m/2+1,\cdots,d\}$, with $\omega\in(0,1)$.\\

\begin{thm}\label{th2}
 $(i)$   All $m\otimes m$ entangled states $\ket{\phi_m}=\sum_{a=1}^m\nu_a\ket{\eta_a}\ket{a},(\nu_a\neq 0, \forall a)$ are useful for the antidiscrimination of three measurements described at \eqref{S} where $m=4p$ and $p\in(1,\cdots,x)$ when $\omega$ takes value such that
    \bea\label{omega_range}
&&\omega^2\leqslant\min_{a\in\{1,\cdots,m\}}\left\{\frac{|\nu_{a+(-1)^{a+1}}|^2}{3|\nu_{a}|^2+|\nu_{a+(-1)^{a+1}}|^2},\right.\nonumber\\
&&\left.\frac{|\nu_{m+1-a}|^2}{3|\nu_{a}|^2+|\nu_{m+1-a}|^2}\right\}.\nonumber\\
    \eea

$(ii)$ The measurements described at \eqref{S} is not antidistinguishable with single system probe.
\end{thm}
Using \eqref{AME}, antidistinguishability of these measurements reduces to the antidistinguishability of reduced states of Bob's side summed over all outcomes. We utilize the sufficient condition for the antidistinguishability of three states given by \cite{barett} and therefore, the sufficient condition of the range of $\omega$ comes. A simple application of Lemma \ref{lem2} shows that single system cannot antidistinguish this set of measurements. For more details, see Appendix \ref{pr2}.

For the rest of the even dimensions $n>4$,
consider the following three projective measurements:
\bea\label{Q}
\{Q_{a|1}\}_{a=1}^n &=& \{\ket{\zeta_a}\bra{\zeta_a}\}_{a=1}^n,
\{Q_{a|2}\}_{a=1}^n = \{\ket{\psi^q_a}\bra{\psi^q_a}\}_{a=1}^n,\nonumber\\
\{Q_{a|3}\}_{a=1}^n &=& \{\ket{\Bar{\psi^q_a}}\bra{\Bar{\psi^q_a}}\}_{a=1}^n,
\eea 
where,
$\ket{\psi^q_a}=\epsilon\ket{\zeta_a}+\sqrt{1-\epsilon^2}\ket{\zeta_{a+1}}$ for $a=\{1,3,5,\cdots,n-1\}$ and $\ket{\psi^q_a}=\sqrt{1-\epsilon^2}\ket{\zeta_{a-1}}-\epsilon\ket{\zeta_{a}}$ for $a=\{2,4,6,\cdots,n\}$,\\
$\ket{\Bar{\psi^q_a}}=\epsilon\ket{\zeta_a}+\sqrt{1-\epsilon^2}\ket{\zeta_{n+1-a}}$ for $a=\{1,2,\cdots,\frac{n-4}{2}\}$, $\ket{\Bar{\psi^q_a}}=-\epsilon\ket{\zeta_a}+\sqrt{1-\epsilon^2}\ket{\zeta_{n+1-a}}$ for $a=\{\frac{n+6}{2},\cdots,n\}$, $\ket{\Bar{\psi^q_a}}=\epsilon\ket{\zeta_a}+\sqrt{1-\epsilon^2}\ket{\zeta_{a+2}}$ for $a=\{\frac{n-2}{2},\frac{n}{2}\}$ and $\ket{\Bar{\psi^q_a}}=\sqrt{1-\epsilon^2}\ket{\zeta_{a-2}}-\epsilon\ket{\zeta_{a}}$ for $a=\{\frac{n+2}{2},\frac{n+4}{2}\}$ with $\epsilon\in (0,1)$.
\begin{thm}
  $(i)$  All $n\otimes n$ entangled states $\ket{\phi_n}=\sum_{a=1}^n\theta_a\ket{\zeta_a}\ket{a}, (\zeta_a\neq 0, \forall a),$ are useful for the antidiscrimination of three measurements given by \eqref{Q}, where $n=4p+2$ and $p\in(1,\cdots,x)$ when $\epsilon$ takes value such that
    \bea\label{epsilon_range}
    &&\epsilon^2\leqslant\min_{a\in\{1,\cdots,n\}}\left\{\frac{|\theta_{a+1}|^2}{3|\theta_a|^2+|\theta_{a+1}|^2}, \frac{|\theta_{n+1-a}|^2}{3|\theta_a|^2+|\theta_{n+1-a}|^2},\right.\nonumber\\
    &&\left.\frac{|\theta_{a+2}|^2}{3|\theta_a|^2+|\theta_{a+2}|^2},
    \frac{|\theta_{a-2}|^2}{3|\theta_a|^2+|\theta_{a-2}|^2}, \frac{|\theta_{a-1}|^2}{3|\theta_a|^2+|\theta_{a-1}|^2}\right\}.\nonumber\\
    \eea
    $(ii)$ The measurements of \eqref{Q} are not antidistinguishable with single system probe.
\end{thm}
The proof uses the same technique as the Theorem \ref{th2}. The proof is deferred to Appendix \ref{pr3}.

\section{Conclusion} This work proves the usefulness all $d\otimes d$ dimensional pure entangled states in the antidiscrimination of quantum measurements when $d$ is even. The measurements taken here are not distinguishable and on the top of that, at each outcome of the measurements, the reduced states are not distinguishable in entanglement assisted protocol. It is easy to see that our construction is not extendable for entangled states of $\mathbbm{C}^d\otimes\mathbbm{C}^d$, when $d$ is odd.

The immediate future problem includes the check of the same statement for odd dimensions too by not giving away the indistinguishability condition. It is not necessary to stick to only measurement channels, one can think of some other kinds of channels where the prowess of all entangled states can be examined in the task of antidiscrimination. This work is of interest not only within the framework of Quantum Information Theory, but also can be useful in the context of modern studies of the complex dynamics of quantum systems. In particular, the antidistinguishability structures can potentially be interpreted in terms of quantum chaos\cite{chirikov1995quantum,casati1989quantum,casati2007dynamical}, dynamic localization, fractal geometry of quantum-state spaces, nonlinear quantum evolution\cite{glushkov2017non,zurek1991decoherence,glushkov2021nonlinear}, entropy and different information measures. 

\subsection*{Acknowledgment} 
The author thanks Debashis Saha and Anandamay Das Bhowmik for fruitful discussion. The author also thanks the anonymous reviewer for the helpful suggestions and insightful comments.

\appendix

\onecolumngrid

\section{Proof of Theorem 1}\label{pr1}
Leveraging the formula for the antidistinguishability of measurements, we can express the antidistinguishability of the three measurements $\{R_{a|x}\}_{x=1}^3$ as

\bea\label{ame_th1}
&& AME\left[\left\{R_{a|x}\right\}_{a,x},\left\{1/3\right\}_x\right]\nonumber\\
&=& \frac13 AS\Bigg[\Bigg\{\ket{1},\frac{\sqrt{\lambda}\cos x\ket{1}+\sqrt{1-\lambda}\sin x\ket{2}}{(\lambda\cos^2 x + (1-\lambda)\sin^2 x)^{1/2}}, \frac{\sqrt{\lambda}\cos x\ket{1}-\sqrt{1-\lambda}\sin x\ket{2}}{(\lambda\cos^2 x + (1-\lambda)\sin^2 x)^{1/2}} \Bigg\},\nonumber\\
&&\quad \Big\{\lambda, \lambda\cos^2 x + (1-\lambda)\sin^2 x, \lambda\cos^2 x + (1-\lambda)\sin^2 x\Big\}\Bigg]\nonumber\\
&& +\frac13 AS\Bigg[\Bigg\{\ket{2}, \frac{\sqrt{\lambda}\sin x\ket{1}-\sqrt{1-\lambda}\cos x\ket{2}}{(\lambda\sin^2 x + (1-\lambda)\cos^2 x)^{1/2}}, \frac{\sqrt{\lambda}\sin x\ket{1}+\sqrt{1-\lambda}\cos x\ket{2}}{(\lambda\sin^2 x + (1-\lambda)\cos^2 x)^{1/2}}\Bigg\},\nonumber\\
&&\quad \Big\{1-\lambda, \lambda\sin^2 x + (1-\lambda)\cos^2 x, \lambda\sin^2 x + (1-\lambda)\cos^2 x\Big\}\Bigg].
\eea

To check the antidistinguishability of these two sets of reduced states, we use the condition of Lemma 1 (proposition $6$ of \cite{Heinosaari_2018}). For the first set, we obtain the following equations:

\bea
&(i)& \mu_1+\mu_2\frac{\lambda \cos^2 x}{(\lambda\cos^2 x + (1-\lambda)\sin^2 x)}+\mu_3\frac{\lambda \cos^2 x}{(\lambda\cos^2 x + (1-\lambda)\sin^2 x)}=1,\\
&(ii)& \mu_2\frac{(1-\lambda) \sin^2 x}{(\lambda\cos^2 x + (1-\lambda)\sin^2 x)}+\mu_3\frac{(1-\lambda) \sin^2 x}{(\lambda\cos^2 x + (1-\lambda)\sin^2 x)}=1,\\
&(iii)& \mu_2\frac{\sqrt{\lambda(1-\lambda)} \cos x\sin x}{(\lambda\cos^2 x + (1-\lambda)\sin^2 x)}-\mu_3\frac{\sqrt{\lambda(1-\lambda)} \cos x\sin x}{(\lambda\cos^2 x + (1-\lambda)\sin^2 x)}=0.
\eea

Solving these, we find

\bea
&& \mu_2=\mu_3=\frac{(\lambda\cos^2 x + (1-\lambda)\sin^2 x)}{2(1-\lambda)\sin^2 x},\\
&& \mu_1=1-\frac{\lambda\cos^2 x}{(1-\lambda)\sin^2 x}.
\eea

All coefficients must be positive. One can check that $\mu_2$ and $\mu_3$ are always positive. For $\mu_1 > 0$, we require

\be\label{cth1}
1-\frac{\lambda\cos^2 x}{(1-\lambda)\sin^2 x} > 0 \quad \Rightarrow \quad \tan^2 x > \frac{\lambda}{1-\lambda}.
\ee

Similarly, for the second set, we obtain

\bea
&(i)& \mu'_1+\mu'_2\frac{(1-\lambda) \cos^2 x}{(\lambda\sin^2 x + (1-\lambda)\cos^2 x)}+\mu'_3\frac{(1-\lambda) \cos^2 x}{(\lambda\sin^2 x + (1-\lambda)\cos^2 x)}=1,\\
&(ii)& \mu'_2\frac{\lambda \sin^2 x}{(\lambda\sin^2 x + (1-\lambda)\cos^2 x)}+\mu'_3\frac{\lambda \sin^2 x}{(\lambda\sin^2 x + (1-\lambda)\cos^2 x)}=1,\\
&(iii)& -\mu'_2\frac{\sqrt{\lambda(1-\lambda)} \cos x\sin x}{(\lambda\sin^2 x + (1-\lambda)\cos^2 x)}+\mu'_3\frac{\sqrt{\lambda(1-\lambda)} \cos x\sin x}{(\lambda\sin^2 x + (1-\lambda)\cos^2 x)}=0.
\eea

Solving these gives

\bea
&& \mu'_2=\mu'_3=\frac{(\lambda\sin^2 x + (1-\lambda)\cos^2 x)}{2\lambda\sin^2 x},\\
&& \mu'_1=1-\frac{(1-\lambda)\cos^2 x}{\lambda\sin^2 x}.
\eea

Imposing $\mu'_1 > 0$ leads to

\be\label{cth2}
\tan^2 x > \frac{1-\lambda}{\lambda}.
\ee

Combining \eqref{cth1} and \eqref{cth2}, the necessary condition is

\be
\tan^2 x > \max\left\{\frac{\lambda}{1-\lambda}, \frac{1-\lambda}{\lambda}\right\}.
\ee

With this condition, both the sets of reduced states are antidistinguishable. So, \eqref{ame_th1} gives,

\bea
&&AME\left[\left\{R_{a|x}\right\}_{a,x},\left\{1/3\right\}_x\right]\nonumber\\
&=&\frac13\left[\lambda+\lambda\cos^2 x + (1-\lambda)\sin^2 x + \lambda\cos^2 x + (1-\lambda)\sin^2 x + 1-\lambda + \lambda\sin^2 x + (1-\lambda)\cos^2 x \right.\nonumber\\
&&+\left.\lambda\sin^2 x + (1-\lambda)\cos^2 x \right]\nonumber\\
&=& 1.
\eea

This completes the proof of our claim that every $2\otimes 2$ entangled state is useful for the antidiscrimination of three measurements. 

As the dimension is two, there exists only one $\sigma$ such that $\tr\left(\sigma R_{a=1|x=x'}\right)=0$. For the same $\sigma$ to satisfy $\tr\left(\sigma R_{a=2|x\neq x'}\right)=0$, the only possibility is $R_{a=1|x=x'}=R_{a=2|x\neq x'}$. Hence, all sets $\{R_{a|x}\}_x$ must share at least one common element $R_{\{a=a^*|x=x'\}}$. 
However, according to the definition of $x$, the sets $\{R_{1|1}, R_{1|2}, R_{1|3}\}$ and $\{R_{2|1}, R_{2|2}, R_{2|3}\}$ do not have any common element. Therefore, we conclude that $AMS < 1$.

\section{Proof of Theorem 2}\label{pr2}
We can write the antidistinguishability of three measurements $\{S_{a|x}\}_{a,x}$ as follows,
    \bea
     && AME\left[\{S_{a|x}\}_{a,x},\{1/3\}_x\right]\nonumber\\
     &=& \frac13 \sum_{a\in\{1,3,5,\cdots,\frac{m}{2}-1\}} AS\left[\left\{\ket{a},\frac{\nu_a\omega\ket{a}+\nu_{a+1}\sqrt{1-\omega^2}\ket{a+1}}{(|\nu_a|^2\omega^2+|\nu_{a+1}|^2(1-\omega^2))^{1/2}},\frac{\nu_a\omega\ket{a}+\nu_{m+1-a}\sqrt{1-\omega^2}\ket{m+1-a}}{(|\nu_a|^2\omega^2+|\nu_{m+1-a}|^2(1-\omega^2))^{1/2}}\right\},\right.\nonumber\\
     &&\left. \left\{|\nu_a|^2, (|\nu_a|^2\omega^2+|\nu_{a+1}|^2(1-\omega^2)), (|\nu_a|^2\omega^2+|\nu_{m+1-a}|^2(1-\omega^2))\right\}\right]\nonumber\\
     &&+\frac13 \sum_{a\in\{\frac{m}{2}+1,\frac{m}{2}+3\cdots,m-1\}}AS\left[\left\{\ket{a},\frac{\nu_a\omega\ket{a}+\nu_{a+1}\sqrt{1-\omega^2}\ket{a+1}}{(|\nu_a|^2\omega^2+|\nu_{a+1}|^2(1-\omega^2)^{1/2}},\frac{\nu_{m+1-a}\sqrt{1-\omega^2}\ket{m+1-a}-\nu_a\omega\ket{a}}{(|\nu_{m+1-a}|^2(1-\omega^2)+|\nu_a|^2\omega^2)^{1/2}}\right\},\right.\nonumber\\
     &&\left. \left\{|\nu_a|^2, (|\nu_a|^2\omega^2+|\nu_{a+1}|^2(1-\omega^2), (|\nu_{m+1-a}|^2(1-\omega^2)+|\nu_a|^2\omega^2)\right\}\right]\nonumber
     \eea
     \bea
     &&+\frac13 \sum_{a\in\{2,4,\cdots,\frac{m}{2}\}} AS\left[\left\{\ket{a},\frac{\nu_{a-1}\sqrt{1-\omega^2}\ket{a-1}-\nu_a\omega\ket{a}}{(|\nu_{a-1}|^2(1-\omega^2)+|\nu_a|^2\omega^2)^{1/2}},\frac{\nu_a\omega\ket{a}+\nu_{m+1-a}\sqrt{1-\omega^2}\ket{m+1-a}}{(|\nu_a|^2\omega^2+|\nu_{m+1-a}|^2(1-\omega^2))^{1/2}}\right\},\right.\nonumber\\
     &&\left. \left\{|\nu_a|^2, (|\nu_{a-1}|^2(1-\omega^2)+|\nu_a|^2\omega^2), (|\nu_a|^2\omega^2+|\nu_{m+1-a}|^2(1-\omega^2))\right\}\right]\nonumber\\
     &&+\frac13 \sum_{a\in\{\frac{m}{2}+2,\frac{m}{2}+4,\cdots,m\}}AS\left[\left\{\ket{a},\frac{\nu_{a-1}\sqrt{1-\omega^2}\ket{a-1}-\nu_a\omega\ket{a}}{(|\nu_{a-1}|^2(1-\omega^2)+|\nu_a|^2\omega^2)^{1/2}},\frac{\nu_{d+1-a}\sqrt{1-\omega^2}\ket{m+1-a}-\nu_a\omega\ket{a}}{(|\nu_{m+1-a}|^2(1-\omega^2)+|\nu_a|^2\omega^2)^{1/2}}\right\},\right.\nonumber\\
     &&\left. \left\{|\nu_a|^2, (|\nu_{a-1}|^2(1-\omega^2)+|\nu_a|^2\omega^2), (|\nu_{m+1-a}|^2(1-\omega^2)+|\nu_a|^2\omega^2)\right\}\right]\nonumber\\
    \eea  
     Now we will use the sufficient condition of antidistinguishability proposed in \cite{barett}. For first set of reduced states for a fixed $a\in\{1,3,5,\cdots,\frac{m}{2}-1\}$, 
    we can write
    \bea
&(i)&\frac{|\nu_a|^2\omega^2}{(|\nu_a|^2\omega^2+|\nu_{a+1}|^2(1-\omega^2))} \leqslant \frac14,\\
&(ii)&\frac{|\nu_a|^2\omega^2}{(|\nu_a|^2\omega^2+|\nu_{m+1-a}|^2(1-\omega^2))} \leqslant \frac14,\\
&(iii)& \frac{|\nu_a|^4\omega^4}{(|\nu_a|^2\omega^2+|\nu_{a+1}|^2(1-\omega^2))(|\nu_a|^2\omega^2+|\nu_{m+1-a}|^2(1-\omega^2))} \leqslant \frac14.
    \eea
The last condition is always true if the first two conditions are. So we will drop the last condition from the calculation and this fact is true for every set. From first two conditions, we arrive at
\bea\label{4_c_11}
&&\omega^2\leqslant\frac{|\nu_{a+1}|^2}{3|\nu_a|^2+|\nu_{a+1}|^2},\\
&&\omega^2\leqslant\frac{|\nu_{m+1-a}|^2}{3|\nu_a|^2+|\nu_{m+1-a}|^2}.\label{4_c_12}
\eea
From the second set with $a\in\{\frac{m}{2}+1,\frac{m}{2}+3\cdots,m-1\}$, we need to get
   \bea
&(i)&\frac{|\nu_a|^2\omega^2}{(|\nu_a|^2\omega^2+|\nu_{a+1}|^2(1-\omega^2))} \leqslant \frac14,\\
&(ii)&\frac{|\nu_{a}|^2\omega^2}{(|\nu_a|^2\omega^2+|\nu_{m+1-a}|^2(1-\omega^2))} \leqslant \frac14.
\eea
These two inequalities give,
\bea\label{4_c_21}
&&\omega^2\leqslant\frac{|\nu_{a+1}|^2}{3|\nu_a|^2+|\nu_{a+1}|^2},\\
&&\omega^2\leqslant\frac{|\nu_{m+1-a}|^2}{3|\nu_a|^2+|\nu_{m+1-a}|^2}.\label{4_c_22}
\eea
Similarly, we obtain for $a\in\{2,4,\cdots,\frac{m}{2}\}$,
\bea
&(i)&\frac{|\nu_a|^2\omega^2}{(|\nu_a|^2\omega^2+|\nu_{a-1}|^2(1-\omega^2))} \leqslant \frac14,\\
&(ii)&\frac{|\nu_a|^2\omega^2}{(|\nu_a|^2\omega^2+|\nu_{m+1-a}|^2(1-\omega^2))} \leqslant \frac14.
\eea
Therefore,
\bea\label{4_c_31}
&&\omega^2\leqslant\frac{|\nu_{a-1}|^2}{3|\nu_a|^2+|\nu_{a-1}|^2},\\
&&\omega^2\leqslant\frac{|\nu_{m+1-a}|^2}{3|\nu_a|^2+|\nu_{m+1-a}|^2}.\label{4_c_32}
\eea
Fourth group of sets give,
\bea
&(i)&\frac{|\nu_a|^2\omega^2}{(|\nu_a|^2\omega^2+|\nu_{a-1}|^2(1-\omega^2))} \leqslant \frac14,\\
&(ii)&\frac{|\nu_a|^2\omega^2}{(|\nu_a|^2\omega^2+|\nu_{m+1-a}|^2(1-\omega^2))} \leqslant \frac14.
\eea
Therefore,
\bea\label{4_c_41}
&&\omega^2\leqslant\frac{|\nu_{a-1}|^2}{3|\nu_a|^2+|\nu_{a-1}|^2},\\
&&\omega^2\leqslant\frac{|\nu_{m+1-a}|^2}{3|\nu_a|^2+|\nu_{m+1-a}|^2}.\label{4_c_42}
\eea
From \eqref{4_c_11},\eqref{4_c_12}, \eqref{4_c_21} and \eqref{4_c_22}, we get sufficient range of $\omega$ for odd $m$, which is,
\bea
\omega^2\leqslant\left\{\frac{|\nu_{a+1}|^2}{3|\nu_a|^2+|\nu_{a+1}|^2},\frac{|\nu_{m+1-a}|^2}{3|\nu_a|^2+|\nu_{m+1-a}|^2}\right\}.
\eea
From \eqref{4_c_31}, \eqref{4_c_32}, \eqref{4_c_41} and \eqref{4_c_42}, we get sufficient value of $\omega$ for even $m$, i.e.,
\bea
\omega^2\leqslant\left\{\frac{|\nu_{a-1}|^2}{3|\nu_a|^2+|\nu_{a+1}|^2},\frac{|\nu_{m+1-a}|^2}{3|\nu_a|^2+|\nu_{m+1-a}|^2}\right\}.
\eea
From this two above conditions, we arrive at
final sufficient range of $\omega$, which is
\bea\label{omega1_range}
&&\omega^2\leqslant\min_{a\in\{1,\cdots,m\}}\left\{\frac{|\nu_{a+(-1)^{a+1}}|^2}{3|\nu_{a}|^2+|\nu_{a+(-1)^{a+1}}|^2}, \frac{|\nu_{m+1-a}|^2}{3|\nu_{a}|^2+|\nu_{m+1-a}|^2}\right\},\nonumber\\
    \eea
for $m=4p$ and $p\in(1,\cdots,x)$.
One can check if the reduced states are antidistinguishable, then,

\bea
&&AME\left[\left\{S_{a|x}\right\}_{a,x},\left\{1/3\right\}_x\right]\nonumber\\
&=&\frac13\left[\sum_{a\in\{1,3,5,\cdots,\frac{m}{2}-1\}}\left(|\nu_a|^2 + |\nu_a|^2\omega^2+|\nu_{a+1}|^2(1-\omega^2) + |\nu_a|^2\omega^2+|\nu_{m+1-a}|^2(1-\omega^2) \right)\right.\nonumber\\
&&+ \sum_{a\in\{\frac{m}{2}+1,\frac{m}{2}+3\cdots,m-1\}} \left(|\nu_a|^2+ |\nu_a|^2\omega^2+|\nu_{a+1}|^2(1-\omega^2) + |\nu_{m+1-a}|^2(1-\omega^2)+|\nu_a|^2\omega^2\right)\nonumber\\
&&+ \sum_{a\in\{2,4,\cdots,\frac{m}{2}\}} \left(|\nu_a|^2 + |\nu_{a-1}|^2(1-\omega^2)+|\nu_a|^2\omega^2 + |\nu_a|^2\omega^2+|\nu_{m+1-a}|^2(1-\omega^2)\right)\nonumber\\
&&+\left. \sum_{a\in\{\frac{m}{2}+2,\frac{m}{2}+4,\cdots,m\}} \left(|\nu_a|^2 + |\nu_{a-1}|^2(1-\omega^2)+|\nu_a|^2\omega^2 + |\nu_{m+1-a}|^2(1-\omega^2)+|\nu_a|^2\omega^2\right)\right]\nonumber\\
&=& 1.
\eea
For AMS, let us take the optimum probe $\Omega$. 
\bea
&& AMS\left[\{S_{a|x}\}_{a,x},\{1/3\}_x\right]\nonumber\\
&=&1-\min_{\Omega}\sum_a\left[\min\left\{|\la\Omega|\eta_a\ra|^2, |\la\Omega|\psi^s_a\ra|^2, |\la\Omega|\Bar{\psi}^s_a\ra|^2\right\}\right].
\eea
At least one of the three terms inside the third bracket needs to be zero for all $a$. As $\Omega$ is $m$-dimensional state, it can be orthogonal to at most $(m-1)$ states. To make AMS $=1$, $\Omega$ needs to be orthogonal to $m$ states. This case is possible only when at least two of the $m$ number of states are same. This is not the case for our measurements discussed here. So the measurements are not antidistinguishable with single system probe. 

\section{Proof of Theorem 3}\label{pr3}
  We can write the antidistinguishability of three measurements $\{Q_{a|x}\}_{a,x}$ as,
    \bea
     && AME\left[\{Q_{a|x}\}_{a,x},\{1/3\}_x\right]\nonumber\\
     &=& \frac13 \sum_{a\in\{1,3,5,\cdots,\frac{n-4}{2}\}} AS\left[\left\{\ket{a},\frac{\theta_a\epsilon\ket{a}+\theta_{a+1}\sqrt{1-\epsilon^2}\ket{a+1}}{(|\theta_a|^2\epsilon^2+|\theta_{a+1}|^2(1-\epsilon^2))^{1/2}},\frac{\theta_a\epsilon\ket{a}+\theta_{n+1-a}\sqrt{1-\epsilon^2}\ket{n+1-a}}{(|\theta_a|^2\epsilon^2+|\theta_{n+1-a}|^2(1-\epsilon^2))^{1/2}}\right\},\right.\nonumber\\
     &&\left. \left\{|\theta_a|^2, (|\theta_a|^2\epsilon^2+|\theta_{a+1}|^2(1-\epsilon^2)), (|\theta_a|^2\epsilon^2+|\theta_{n+1-a}|^2(1-\epsilon^2))\right\}\right]\nonumber
     \eea
     \bea
     &&+\frac13 \sum_{a\in\{\frac{n+6}{2}+1,\frac{n+6}{2}+3\cdots,n-1\}}AS\left[\left\{\ket{a},\frac{\theta_a\epsilon\ket{a}+\theta_{a+1}\sqrt{1-\epsilon^2}\ket{a+1}}{(|\theta_a|^2\epsilon^2+|\theta_{a+1}|^2(1-\epsilon^2)^{1/2}},\frac{\theta_{n+1-a}\sqrt{1-\epsilon^2}\ket{n+1-a}-\theta_a\epsilon\ket{a}}{(|\theta_{n+1-a}|^2(1-\epsilon^2)+|\theta_a|^2\epsilon^2)^{1/2}}\right\},\right.\nonumber\\
     &&\left. \left\{|\theta_a|^2, (|\theta_a|^2\epsilon^2+|\theta_{a+1}|^2(1-\epsilon^2), (|\theta_{n+1-a}|^2(1-\epsilon^2)+|\theta_a|^2\epsilon^2)\right\}\right]\nonumber\\
     &&+\frac13 \sum_{a\in\{\frac{n}{2}\}} AS\left[\left\{\ket{a},\frac{\theta_a\epsilon\ket{a}+\theta_{a+1}\sqrt{1-\epsilon^2}\ket{a+1}}{(|\theta_a|^2\epsilon^2+|\theta_{a+1}|^2(1-\epsilon^2))^{1/2}},\frac{\theta_a\epsilon\ket{a}+\theta_{a+2}\sqrt{1-\epsilon^2}\ket{a+2}}{(|\theta_a|^2\epsilon^2+|\theta_{a+2}|^2(1-\epsilon^2))^{1/2}}\right\},\right.\nonumber\\
     &&\left. \left\{|\theta_a|^2, (|\theta_a|^2\epsilon^2+|\theta_{a+1}|^2(1-\epsilon^2)), (|\theta_a|^2\epsilon^2+|\theta_{a+2}|^2(1-\epsilon^2))\right\}\right]\nonumber\\
     &&+\frac13 \sum_{a\in\{\frac{n+4}{2}\}}AS\left[\left\{\ket{a},\frac{\theta_a\epsilon\ket{a}+\theta_{a+1}\sqrt{1-\epsilon^2}\ket{a+1}}{(|\theta_a|^2\epsilon^2+|\theta_{a+1}|^2(1-\epsilon^2))^{1/2}},\frac{\theta_{a-2}\sqrt{1-\epsilon^2}\ket{a-2}-\theta_a\epsilon\ket{a}}{(|\theta_{a-2}|^2(1-\epsilon^2)+|\theta_a|^2\epsilon^2)^{1/2}}\right\},\right.\nonumber\\
     &&\left. \left\{|\theta_a|^2, (|\theta_a|^2\epsilon^2+|\theta_{a+1}|^2(1-\epsilon^2)), (|\theta_{a-2}|^2(1-\epsilon^2)+|\theta_a|^2\epsilon^2)\right\}\right]\nonumber\\
     &&+ \frac13 \sum_{a\in\{2,4,6,\cdots,\frac{n-4}{2}-1\}} AS\left[\left\{\ket{a},\frac{\theta_{a-1}\sqrt{1-\epsilon^2}\ket{a-1}-\theta_a\epsilon\ket{a}}{(|\theta_{a-1}|^2(1-\epsilon^2)+|\theta_a|^2\epsilon^2)^{1/2}},\frac{\theta_a\epsilon\ket{a}+\theta_{n+1-a}\sqrt{1-\epsilon^2}\ket{n+1-a}}{(|\theta_a|^2\epsilon^2+|\theta_{n+1-a}|^2(1-\epsilon^2))^{1/2}}\right\},\right.\nonumber\\
     &&\left. \left\{|\theta_a|^2, (|\theta_a|^2\epsilon^2+|\theta_{a+1}|^2(1-\epsilon^2)), (|\theta_a|^2\epsilon^2+|\theta_{n+1-a}|^2(1-\epsilon^2))\right\}\right]\nonumber\\
     &&+\frac13 \sum_{a\in\{\frac{n+6}{2},\frac{n+6}{2}+2\cdots,n\}}AS\left[\left\{\ket{a},\frac{\theta_{a-1}\sqrt{1-\epsilon^2}\ket{a-1}-\theta_a\epsilon\ket{a}}{(|\theta_{a-1}|^2(1-\epsilon^2)+|\theta_a|^2\epsilon^2)^{1/2}},\frac{\theta_{n+1-a}\sqrt{1-\epsilon^2}\ket{n+1-a}-\theta_a\epsilon\ket{a}}{(|\theta_{n+1-a}|^2(1-\epsilon^2)+|\theta_a|^2\epsilon^2)^{1/2}}\right\},\right.\nonumber\\
     &&\left. \left\{|\theta_a|^2, (|\theta_a|^2\epsilon^2+|\theta_{a+1}|^2(1-\epsilon^2), (|\theta_{n+1-a}|^2(1-\epsilon^2)+|\theta_a|^2\epsilon^2)\right\}\right]\nonumber
     \eea
     \bea
     &&+\frac13 \sum_{a\in\{\frac{n-2}{2}\}} AS\left[\left\{\ket{a},\frac{\theta_{a-1}\sqrt{1-\epsilon^2}\ket{a-1}-\theta_a\epsilon\ket{a}}{(|\theta_{a-1}|^2(1-\epsilon^2)+|\theta_a|^2\epsilon^2)^{1/2}},\frac{\theta_a\epsilon\ket{a}+\theta_{a+2}\sqrt{1-\epsilon^2}\ket{a+2}}{(|\theta_a|^2\epsilon^2+|\theta_{a+2}|^2(1-\epsilon^2))^{1/2}}\right\},\right.\nonumber\\
     &&\left. \left\{|\theta_a|^2, (|\theta_a|^2\epsilon^2+|\theta_{a+1}|^2(1-\epsilon^2)), (|\theta_a|^2\epsilon^2+|\theta_{a+2}|^2(1-\epsilon^2))\right\}\right]\nonumber\\
     &&+\frac13 \sum_{a\in\{\frac{n+2}{2}\}}AS\left[\left\{\ket{a},\frac{\theta_{a-1}\sqrt{1-\epsilon^2}\ket{a-1}-\theta_a\epsilon\ket{a}}{(|\theta_{a-1}|^2(1-\epsilon^2)+|\theta_a|^2\epsilon^2)^{1/2}},\frac{\theta_{a-2}\sqrt{1-\epsilon^2}\ket{a-2}-\theta_a\epsilon\ket{a}}{(|\theta_{a-2}|^2(1-\epsilon^2)+|\theta_a|^2\epsilon^2)^{1/2}}\right\},\right.\nonumber\\
     &&\left. \left\{|\theta_a|^2, (|\theta_a|^2\epsilon^2+|\theta_{a+1}|^2(1-\epsilon^2)), (|\theta_{a-2}|^2(1-\epsilon^2)+|\theta_a|^2\epsilon^2)\right\}\right]\nonumber\\
    \eea  
     Now we will use the same sufficient condition of antidistinguishability as our previous proof. For first set of reduced states for a fixed $a\in\{1,3,5,\cdots,\frac{n-4}{2}\}$, 
    we can write
    \bea
&(i)&\frac{|\theta_a|^2\epsilon^2}{(|\theta_a|^2\epsilon^2+|\theta_{a+1}|^2(1-\epsilon^2))} \leqslant \frac14\\
&(ii)&\frac{|\theta_a|^2\epsilon^2}{(|\theta_a|^2\epsilon^2+|\theta_{n+1-a}|^2(1-\epsilon^2))} \leqslant \frac14\\
&(iii)& \frac{|\theta_a|^4\epsilon^4}{(|\theta_a|^2\epsilon^2+|\theta_{a+1}|^2(1-\epsilon^2))(|\theta_a|^2\epsilon^2+|\theta_{n+1-a}|^2(1-\epsilon^2))} \leqslant \frac14
    \eea
In the similar fashion to the last result, the last condition is always true if the first two conditions are. From first two conditions, we arrive at
\bea\label{4_d_1}
&&\epsilon^2\leqslant\frac{|\theta_{a+1}|^2}{3|\theta_a|^2+|\theta_{a+1}|^2}\\
&&\epsilon^2\leqslant\frac{|\theta_{n+1-a}|^2}{3|\theta_a|^2+|\theta_{n+1-a}|^2}
\eea
From the second set with $a\in\{\frac{n+6}{2}+1,\frac{n+6}{2}+3\cdots,n-1\}$, we need to get
   \bea
&(i)&\frac{|\theta_a|^2\epsilon^2}{(|\theta_a|^2\epsilon^2+|\theta_{a+1}|^2(1-\epsilon^2))} \leqslant \frac14\\
&(ii)&\frac{|\theta_{a}|^2\epsilon^2}{(|\theta_a|^2\epsilon^2+|\theta_{n+1-a}|^2(1-\epsilon^2))} \leqslant \frac14
\eea
These two inequalities reduce to,
\bea\label{4_d_2}
&&\epsilon^2\leqslant\frac{|\theta_{a+1}|^2}{3|\theta_a|^2+|\theta_{a+1}|^2},\\
&&\epsilon^2\leqslant\frac{|\theta_{n+1-a}|^2}{3|\theta_a|^2+|\theta_{n+1-a}|^2}.
\eea
Similarly, we obtain for $a\in\{\frac{n}{2}\}$,
\bea
&(i)&\frac{|\theta_a|^2\epsilon^2}{(|\theta_a|^2\epsilon^2+|\theta_{a+1}|^2(1-\epsilon^2))} \leqslant \frac14\\
&(ii)&\frac{|\theta_a|^2\epsilon^2}{(|\theta_a|^2\epsilon^2+|\theta_{a+2}|^2(1-\epsilon^2))} \leqslant \frac14.
\eea
It follows that,
\bea\label{4_c_3}
&&\epsilon^2\leqslant\frac{|\theta_{a+1}|^2}{3|\theta_a|^2+|\theta_{a+1}|^2},\\
&&\epsilon^2\leqslant\frac{|\theta_{a+2}|^2}{3|\theta_a|^2+|\theta_{a+2}|^2}.
\eea
Fourth group $a\in\{\frac{n+4}{2}\}$ gives,
\bea
&(i)&\frac{|\theta_a|^2\epsilon^2}{(|\theta_a|^2\epsilon^2+|\theta_{a+1}|^2(1-\epsilon^2))} \leqslant \frac14\\
&(ii)&\frac{|\theta_a|^2\epsilon^2}{(|\theta_a|^2\epsilon^2+|\theta_{a-2}|^2(1-\epsilon^2))} \leqslant \frac14.
\eea
Therefore,
\bea\label{4_c_4}
&&\epsilon^2\leqslant\frac{|\theta_{a+1}|^2}{3|\theta_a|^2+|\theta_{a+1}|^2},\\
&&\epsilon^2\leqslant\frac{|\theta_{a-2}|^2}{3|\theta_a|^2+|\theta_{a-2}|^2}.
\eea
From the fifth set with $a\in\{2,4,6,\cdots,\frac{n-4}{2}-1\}$, we have
   \bea
&(i)&\frac{|\theta_a|^2\epsilon^2}{(|\theta_a|^2\epsilon^2+|\theta_{a-1}|^2(1-\epsilon^2))} \leqslant \frac14\\
&(ii)&\frac{|\theta_{a}|^2\epsilon^2}{(|\theta_a|^2\epsilon^2+|\theta_{n+1-a}|^2(1-\epsilon^2))} \leqslant \frac14
\eea
From above two inequalities, we obtain
\bea\label{4_d_2}
&&\epsilon^2\leqslant\frac{|\theta_{a-1}|^2}{3|\theta_a|^2+|\theta_{a-1}|^2},\\
&&\epsilon^2\leqslant\frac{|\theta_{n+1-a}|^2}{3|\theta_a|^2+|\theta_{n+1-a}|^2}.
\eea
For $a\in\{\frac{n+6}{2},\frac{n+6}{2}+2\cdots,n\}$,
\bea
&(i)&\frac{|\theta_a|^2\epsilon^2}{(|\theta_a|^2\epsilon^2+|\theta_{a+1}|^2(1-\epsilon^2))} \leqslant \frac14\\
&(ii)&\frac{|\theta_a|^2\epsilon^2}{(|\theta_a|^2\epsilon^2+|\theta_{a+2}|^2(1-\epsilon^2))} \leqslant \frac14.
\eea
Thus,
\bea\label{4_c_3}
&&\epsilon^2\leqslant\frac{|\theta_{a+1}|^2}{3|\theta_a|^2+|\theta_{a+1}|^2},\\
&&\epsilon^2\leqslant\frac{|\theta_{a+2}|^2}{3|\theta_a|^2+|\theta_{a+2}|^2}.
\eea
Then $a\in\{\frac{n-2}{2}\}$ provides,
\bea
&(i)&\frac{|\theta_a|^2\epsilon^2}{(|\theta_a|^2\epsilon^2+|\theta_{a+1}|^2(1-\epsilon^2))} \leqslant \frac14\\
&(ii)&\frac{|\theta_a|^2\epsilon^2}{(|\theta_a|^2\epsilon^2+|\theta_{a-2}|^2(1-\epsilon^2))} \leqslant \frac14.
\eea
Therefore,
\bea\label{4_c_4}
&&\epsilon^2\leqslant\frac{|\theta_{a+1}|^2}{3|\theta_a|^2+|\theta_{a+1}|^2},\\
&&\epsilon^2\leqslant\frac{|\theta_{a-2}|^2}{3|\theta_a|^2+|\theta_{a-2}|^2}.
\eea
 $a\in\{\frac{n+2}{2}\}$ delivers,
\bea
&(i)&\frac{|\theta_a|^2\epsilon^2}{(|\theta_a|^2\epsilon^2+|\theta_{a+1}|^2(1-\epsilon^2))} \leqslant \frac14\\
&(ii)&\frac{|\theta_a|^2\epsilon^2}{(|\theta_a|^2\epsilon^2+|\theta_{a-2}|^2(1-\epsilon^2))} \leqslant \frac14.
\eea
Hence,
\bea\label{4_c_4}
&&\epsilon^2\leqslant\frac{|\theta_{a+1}|^2}{3|\theta_a|^2+|\theta_{a+1}|^2},\\
&&\epsilon^2\leqslant\frac{|\theta_{a-2}|^2}{3|\theta_a|^2+|\theta_{a-2}|^2}.
\eea
From the above conditions, we arrive at sufficient range of $\epsilon$, which is,
\bea
\epsilon^2\leqslant\min_{a\in\{1,\cdots,n\}}\left\{\frac{|\theta_{a+1}|^2}{3|\theta_a|^2+|\theta_{a+1}|^2}, \frac{|\theta_{n+1-a}|^2}{3|\theta_a|^2+|\theta_{n+1-a}|^2},\frac{|\theta_{a+2}|^2}{3|\theta_a|^2+|\theta_{a+2}|^2},\frac{|\theta_{a-2}|^2}{3|\theta_a|^2+|\theta_{a-2}|^2}, \frac{|\theta_{a-1}|^2}{3|\theta_a|^2+|\theta_{a-1}|^2}\right\}.
\eea
In this range of $\epsilon$, the reduced states are antidistinguishable, which implies,
\bea
&&AME\left[\left\{Q_{a|x}\right\}_{a,x},\left\{1/3\right\}_x\right]\nonumber\\
&=&\frac13\left[\sum_{a\in\{1,3,5,\cdots,\frac{n-4}{2}\}}\left(|\theta_a|^2 + |\theta_a|^2\epsilon^2+|\theta_{a+1}|^2(1-\epsilon^2) + |\theta_a|^2\epsilon^2+|\theta_{n+1-a}|^2(1-\epsilon^2) \right)\right.\nonumber\\
&&+ \sum_{a\in\{\frac{n+6}{2}+1,\frac{n+6}{2}+3\cdots,n-1\}} \left(|\theta_a|^2+ |\theta_a|^2\epsilon^2+|\theta_{a+1}|^2(1-\epsilon^2) + |\theta_{n+1-a}|^2(1-\epsilon^2)+|\theta_a|^2\epsilon^2\right)\nonumber\\
&&+\sum_{a\in\{\frac{n}{2}\}}\left(|\theta_a|^2 + |\theta_a|^2\epsilon^2+|\theta_{a+1}|^2(1-\epsilon^2) + |\theta_a|^2\epsilon^2+|\theta_{a+2}|^2(1-\epsilon^2) \right)\nonumber\\
&&+ \sum_{a\in\{\frac{n+4}{2}\}} \left(|\theta_a|^2+ |\theta_a|^2\epsilon^2+|\theta_{a+1}|^2(1-\epsilon^2) + |\theta_{a-2}|^2(1-\epsilon^2)+|\theta_a|^2\epsilon^2\right)\nonumber\\
&&+ \sum_{a\in\{2,4,\cdots,\frac{n-4}{2}-1\}} \left(|\theta_a|^2 + |\theta_{a+1}|^2(1-\epsilon^2)+|\theta_a|^2\epsilon^2 + |\theta_a|^2\epsilon^2+|\theta_{n+1-a}|^2(1-\epsilon^2)\right)\nonumber\\
&&+ \sum_{a\in\{\frac{n+6}{2},\frac{n+6}{2}+2,\cdots,n\}} \left(|\theta_a|^2 + |\theta_{a+1}|^2(1-\epsilon^2)+|\theta_a|^2\epsilon^2 + |\theta_{n+1-a}|^2(1-\epsilon^2)+|\theta_a|^2\epsilon^2\right)\nonumber
\eea
\bea
&&+\sum_{a\in\{\frac{n-2}{2}\}}\left(|\theta_a|^2 + |\theta_a|^2\epsilon^2+|\theta_{a+1}|^2(1-\epsilon^2) + |\theta_a|^2\epsilon^2+|\theta_{a+2}|^2(1-\epsilon^2) \right)\nonumber\\
&&+ \left.\sum_{a\in\{\frac{n+2}{2}\}} \left(|\theta_a|^2+ |\theta_a|^2\epsilon^2+|\theta_{a+1}|^2(1-\epsilon^2) + |\theta_{a-2}|^2(1-\epsilon^2)+|\theta_a|^2\epsilon^2\right)\right]\nonumber\\
&=& 1.
\eea

For AMS, let us take the optimum probe $\Sigma$. 
\bea
&& AMS\left[\{Q_{a|x}\}_{a,x},\{1/3\}_x\right]\nonumber\\
&=&1-\min_{\epsilon}\sum_a\left[\min\left\{|\la\Sigma|\eta_a\ra|^2, |\la\Sigma|\psi^q_a\ra|^2, |\la\Sigma|\Bar{\psi}^q_a\ra|^2\right\}\right].
\eea
 As $\Sigma$ is $n$-dimensional state, it can be orthogonal to at most $(n-1)$ states. To make AMS $=1$, $\Sigma$ needs to be orthogonal to a set of states with cardinality $n$. This case is possible only when at least two of the $n$ number of states are same. This is not the case for our defined measurements. So the measurements are not antidistinguishable with single system probe.
\twocolumngrid
\bibliography{ref} 

\end{document}